\documentclass{vgtc}                 

\graphicspath{{figures/}{pictures/}{images/}{./}} 
\usepackage{amssymb}
\usepackage{times}                     
\usepackage{amsmath}
\usepackage{times}   
\usepackage{graphicx}
\usepackage{algpseudocode}
\usepackage[linesnumbered,ruled,vlined]{algorithm2e}  
\usepackage{float} 

\usepackage{multirow}  
\usepackage{graphicx}  
\usepackage{amsmath}   
\usepackage{cite}      

\usepackage{tabu}                      
\usepackage{booktabs}                  
\usepackage{lipsum} 
\usepackage{comment}
\usepackage{mwe}                       
\usepackage{amsmath}  
\usepackage{boldline} 
\usepackage{mathptmx}                  

\usepackage{tabularx} 
\usepackage{array}    

\onlineid{}

\vgtccategory{Research}

\vgtcinsertpkg

\title{SketchPlay: Intuitive Creation of Physically Realistic VR Content with Gesture-Driven Sketching}

\author{
    Xiangwen Zhang\thanks{Equal contribution. e-mail: xiangwen\_zhang@yeah.net} \\
    \scriptsize Beijing Technology and Business University
\and 
    Xiaowei Dai\thanks{Equal contribution. e-mail: 2330702014@st.btbu.edu.cn} \\
    \scriptsize Beijing Technology and Business University
\and 
    Runnan Chen\thanks{e-mail: runnan.chen@sydney.edu.au} \\
    \scriptsize The University of Sydney
\and 
    Xiaoming Chen\thanks{e-mail: xiaoming.chen@btbu.edu.cn} \\
    \scriptsize Beijing Technology and Business University
\and 
    Zeke Zexi Hu\thanks{e-mail: zexi.hu@sydney.edu.au} \\
    \scriptsize The University of Sydney   
}

\teaser{
  \centering
  \includegraphics[width=1\linewidth]{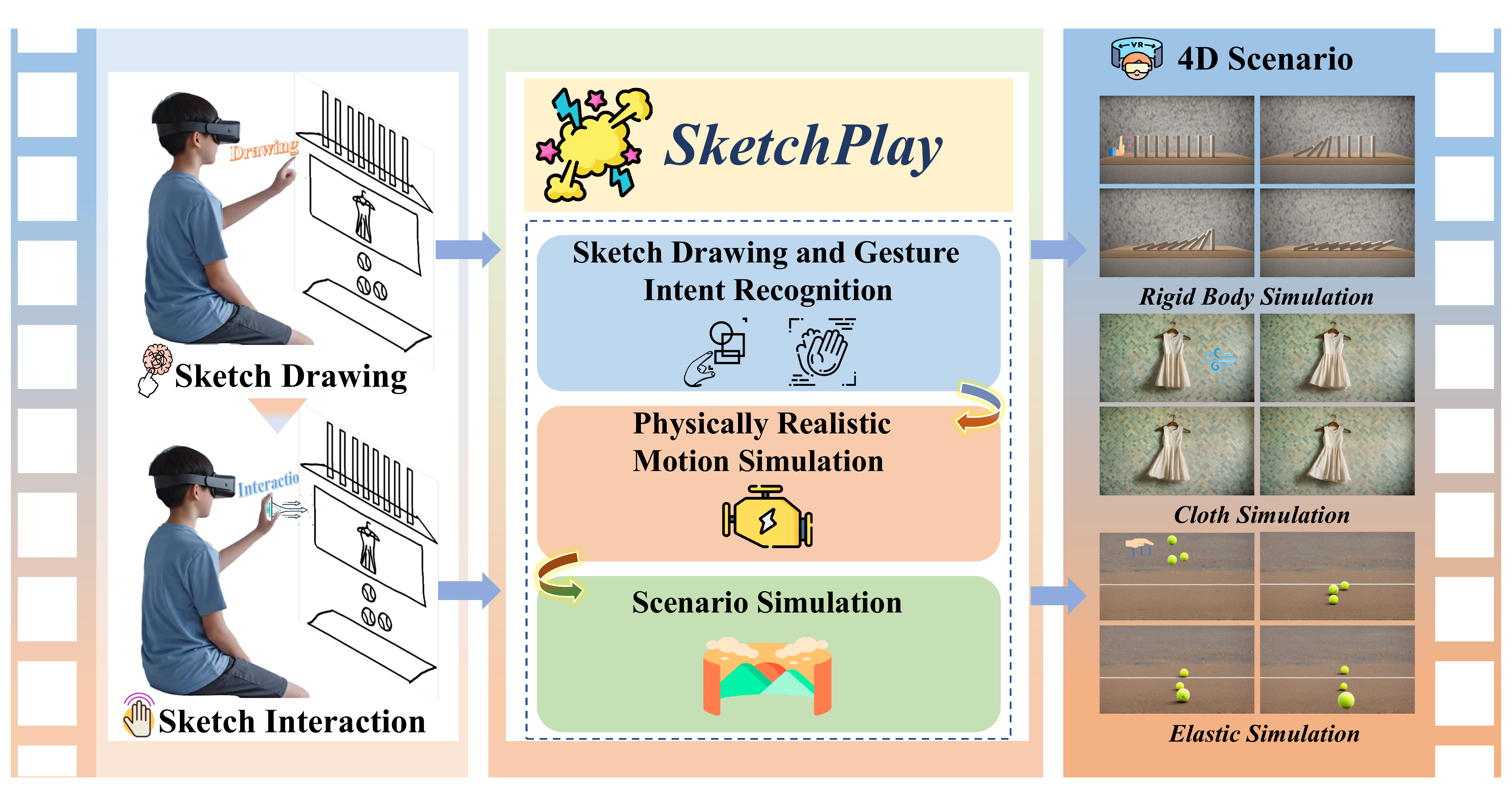}
  \caption{
  \textbf{Overview of SketchPlay. }Existing Virtual Reality (VR) creation tools are commonly constrained by physical controllers, which not only limit a user's interaction freedom but also struggle to transform static sketches into vivid, dynamic scenes. In this paper, we propose \textbf{SketchPlay}, a pipeline that directly converts bare-hand creations into dynamic realities without accessing controllers. Specifically, a user first sketches a static model in mid-air using natural hand gestures. Subsequently, the user can intuitively express dynamic intent through gestures, injecting motion trajectories into the static creation. Finally, SketchPlay captures the user's complete creative intent and automatically synthesizes it into a lifelike, dynamic and physically realistic 4D scene .}
  \label{fig:teaser}
}

\abstract{
Creating physically realistic content in VR often requires complex modeling tools or predefined 3D models, textures, and animations, which present significant barriers for non-expert users. In this paper, we propose SketchPlay, a novel VR interaction framework that transforms humans' air-drawn sketches and gestures into dynamic, physically realistic scenes, making content creation intuitive and playful like drawing. Specifically, sketches capture the structure and spatial arrangement of objects and scenes, while gestures convey physical cues such as velocity, direction, and force that define movement and behavior. By combining these complementary forms of input, SketchPlay captures both the structure and dynamics of user-created content, enabling the generation of a wide range of complex physical phenomena, such as rigid body motion, elastic deformation, and cloth dynamics. Experimental results demonstrate that, compared to traditional text-driven methods, SketchPlay offers significant advantages in expressiveness, and user experience. By providing an intuitive and engaging creation process, SketchPlay lowers the entry barrier for non-expert users and shows strong potential for applications in education, art, and immersive storytelling.
}


\begin{document}



\maketitle

\section{INTRODUCTION}
Virtual reality (VR) has rapidly evolved into a powerful medium for immersive storytelling, creative expression, and interactive simulations. Generating physically realistic content is crucial for achieving immersive and believable experiences, as physical plausibility can enhance visual fidelity and strengthen the user's perception of agency, presence, and interactivity. For example, objects that move, deform, or collide according to physical laws make virtual scenes feel more natural and engaging, which is essential for applications ranging from education and training to art and entertainment.

However, the creation of physically realistic content in VR remains highly challenging for non-expert users. Current solutions for generating physical phenomena, such as rigid body motion, elastic deformation, or cloth simulation, typically require complex modeling tools or predefined 3D assets \cite{faeth2008cutting}. These tools involve steep learning curves and significant operational complexity, creating a barrier to entry for casual creators and hindering the accessibility of VR as a creative medium. This raises a critical problem: \textit{How to design a user-friendly system that can create free-form 4D physically realistic content?}

A gesture-based interface is an intuitive alternative to facilitating natural and seamless interaction between users and computers. It is widely adopted for symbolic tasks like menu navigation or object manipulation \cite{vanichvoranun2024estatig,ying2024enhancing,zhou2024digital}, but it remains underexplored for physics-driven content creation. Besides, existing VR painting and sketching applications \cite{gao2023vr,jiang2021handpainter,thoravi2019tutorivr,suzuki2020realitysketch} illustrate the potential of embodied creativity but still rely heavily on hand-held controllers or wearable devices. Such hardware introduces an artificial layer between the user's body and the virtual space, making interaction indirect and less immersive. Moreover, these approaches often neglect the integration of natural physical cues, such as velocity, direction, and force of a gesture, into the generated scene. As a result, the final content lacks realism and dynamic expressiveness, limiting both the fidelity of user intent and the perceptual impact of the experience.

To overcome these limitations, we present \textbf{SketchPlay}, a novel VR interaction framework that transforms humans' mid-air sketches and gestures into dynamic, physically realistic 4D scenes. The workflow proceeds in three stages. Stage 1 (Hand-drawn Sketch): users first sketch objects in the air, then perform gestures to specify the motion direction. We refine rough sketches with AirSketch\cite{lim2024airsketch}, segment objects, and capture gesture trajectories with MediaPipe \cite{zhang2020mediapipe}, extracting velocity from hand coordinates over time. Stage 2 (Physically Realistic Motion Simulation): sketches, trajectories, and optional text prompts are passed to a vision-language model (VLM) to infer physical attributes such as mass, elasticity, and friction, and generate Blender \cite{Blender} scripts for physics simulation. To improve quality, we curate real-world motion videos and synthetic scripts to form training pairs, and adopt a few-shot prompting strategy to guide the VLM, while SketchDream\cite{liu2024sketchdream} converts sketches into 3D objects. Stage 3 (Scenario Synthesis): Blender outputs edge and depth maps that GPT4Motion\cite{lv2024gpt4motion} processes to produce videos, enhanced by VEnhancer\cite{he2024venhancer} for higher temporal and spatial resolution, expanded to multi-view with ReCamMaster\cite{bai2025recammaster}, and finally reconstructed into coherent 4D scenes with Gaussian Splatting\cite{wu20244d}, ensuring temporal and spatial consistency.

Experiments on quantitative evaluations and qualitative user studies, demonstrate that compared to traditional text-driven methods, SketchPlay achieves superior expressiveness, visual realism, and user experience. By providing an intuitive and engaging creation process, SketchPlay significantly lowers the entry barrier for non-expert users and shows strong potential for applications in education, art, and immersive storytelling.

In summary, this paper makes the following contributions:
\begin{itemize}
\item We propose SketchPlay, a VR system that integrates mid-air sketching with gesture-based interaction to enable intuitive, physics-driven scene creation.
\item We design a two-stage input paradigm—static sketching for structure and dynamic gestures for motion—and leverage a vision-language model to translate these multimodal inputs into executable Blender scripts for physics simulation.
\item We conduct experiments demonstrating that SketchPlay produces scenes with higher expressiveness, visual realism, and physical plausibility compared to traditional controller-based or text-driven methods.
\end{itemize}

\section{Related Work}

\subsection{Gesture-Based Interaction in Virtual Reality}
Gestures have long been recognized as a natural modality for human-computer interaction (HCI). In VR, advances in hand tracking and motion capture technologies have enabled users to interact with virtual environments through pointing, grabbing, and symbolic gestures \cite{laviola20173d, yang2019gesture}. Early work primarily used gestures for menu navigation and object manipulation, emphasizing efficiency and intuitiveness compared to controller-based inputs \cite{piumsomboon2013user, coburn2018effectiveness}. For example, Hürst and Van Wezel \cite{hurst2013gesture} explored gesture-based interaction via finger tracking for mobile augmented reality (AR), demonstrating its potential in creating more immersive experiences in mobile AR applications. More recently, gesture-based systems have been extended to domains such as artistic creation \cite{deepa2025gesture}, collaborative VR \cite{wu2019understanding}, and immersive storytelling \cite{kang2017storytelling}. Similarly, Rautaray and Agrawal \cite{rautaray2011interaction} proposed a hand gesture recognition system for virtual games, highlighting its potential for enhancing user interaction with virtual objects. Kim and Lee \cite{kim2016touch} furthered this exploration by investigating touch and hand gesture-based interactions for mobile AR, proposing a hybrid approach that combines both interaction methods to manipulate 3D objects directly. However, these approaches often treat gestures as \emph{discrete commands} or symbolic triggers, rather than as continuous, physically meaningful inputs. By contrast, SketchPlay interprets both the form of a sketch and the dynamics of the user’s hand motion, using them to drive physically realistic simulations.


\subsection{Sketch-Driven Modeling and Animation}
Sketch-based interfaces provide an intuitive pathway for non-experts to create digital content. From early systems like Sketchpad\cite{sutherland1964sketch} that converted strokes into vector graphics, to later CAD tools that generated 3D shapes from sketches\cite{olsen2009sketch}, this field has consistently aimed to lower the barrier to creation. More recent systems have explored sketch-based animation, enabling the generation of motion paths, poses, and visual effects from drawings \cite{samavati2011sketch, xu2013sketch2scene, mao2005sketch}. Building on this, recent research has further extended these applications with generative models. For instance, SketchDream\cite{liu2024sketchdream} can generate static 3D models from sketches, while state-of-the-art works like SketchVideo\cite{liu2025sketchvideo}, VidSketch\cite{jiang2025vidsketch}, and SparseCtrl\cite{guo2024sparsectrl} utilize sketches as control signals to guide diffusion models in generating or editing dynamic videos, achieving fine-grained control over object shape and motion trajectories.

However, despite the great success of these methods in enhancing visual expressiveness and control precision, they share a common characteristic: the pursuit of visual plausibility rather than physical realism. In these systems, the motion of objects is based on data-driven visual patterns, and their behavior does not adhere to real physical constraints such as mass, elasticity, or collision. To bridge this gap, our SketchPlay takes a novel approach by deeply integrating user gesture sketches directly with a physics engine for the first time. It moves beyond merely reproducing visual appearance, instead capturing the dynamic information of the gesture itself to simulate physically correct and intuitive interactions, thereby achieving true physical realism in the generated scenes.


\subsection{Physics-Based Simulation in Interactive Systems}

Physics engines are widely adopted in virtual reality (VR) and gaming to simulate dynamic phenomena such as rigid bodies, fluids, cloth, and deformable materials, achieving high realism through modeling physical laws like gravity, collision, and friction \cite{millington2007game, bridson2015fluid}. For instance, engines like Unity’s PhysX and Unreal Engine’s Chaos Physics enable realistic interactions in games and VR applications, such as destructible environments and fluid dynamics. These systems excel in entertainment, medical training, and engineering prototyping but require extensive parameter tuning, scripting, or asset design, limiting accessibility to expert users and imposing a steep learning curve. Interactive environments like Algodoo \cite{gregorcic2017algodoo} and Crayon Physics Deluxe simplify physics authoring through intuitive interfaces, enabling sandbox-style experimentation where users draw objects to explore concepts like momentum conservation. Similarly, educational VR platforms like Labster leverage physics-based interactions to illustrate fundamental concepts, such as circuits or molecular dynamics, with real-time feedback  \cite{radianti2020systematic}. Recent advancements, including open-source engines like Bullet Physics integrated with VR headsets, support collaborative simulations \cite{izadi2018simulating}. However, these approaches often rely on preconfigured templates or operate in 2D, limiting creative flexibility in 3D VR environments. SketchPlay addresses this gap by providing a gesture-to-physics interface, using MediaPipe\cite{zhang2020mediapipe} to capture kinematic information (e.g., velocity, direction) from hand-drawn sketches and GPT-4o to infer physical properties (e.g., elasticity, mass), generating complex behaviors from rigid bodies to cloth. This lowers the barrier for non-experts, enhances expressiveness, and bridges intuitive interaction with professional-grade simulation.


\begin{figure*}[t]
  \centering
  \includegraphics[width=\textwidth]{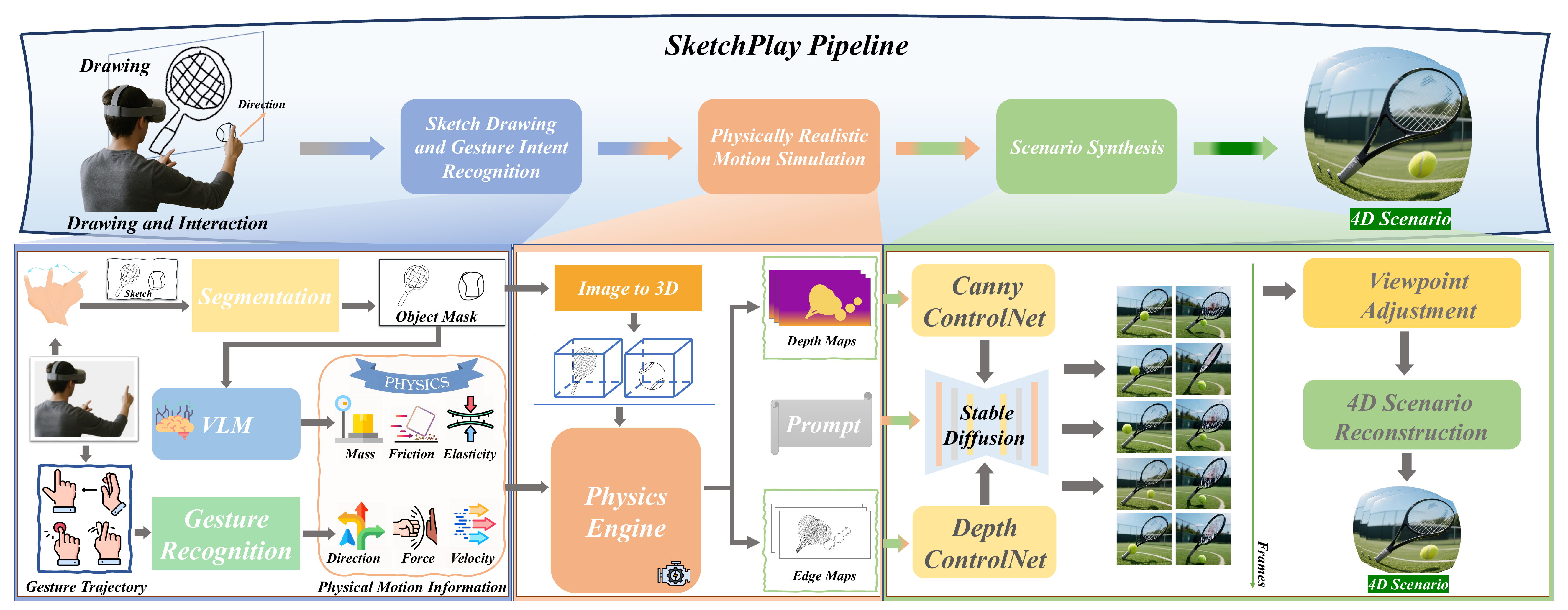} 
  \caption{\textbf{SketchPlay Pipeline}. Our pipeline consists of three core stages: (1) \textbf{Sketch Drawing and Gesture Intent Recognition}, where dynamic physical information like velocity and direction is extracted from the user's gesture; (2) \textbf{Physically Realistic Motion Simulation}, where a physics engine uses this information to simulate realistic behaviors; and (3) \textbf{Scenario Synthesis}, which uses physical priors like edge and depth maps from the simulation to render a photorealistic video.}
  \label{fig:framework2}
\end{figure*}
\begin{figure}[t]
\centering
  \includegraphics[width=0.47\textwidth]{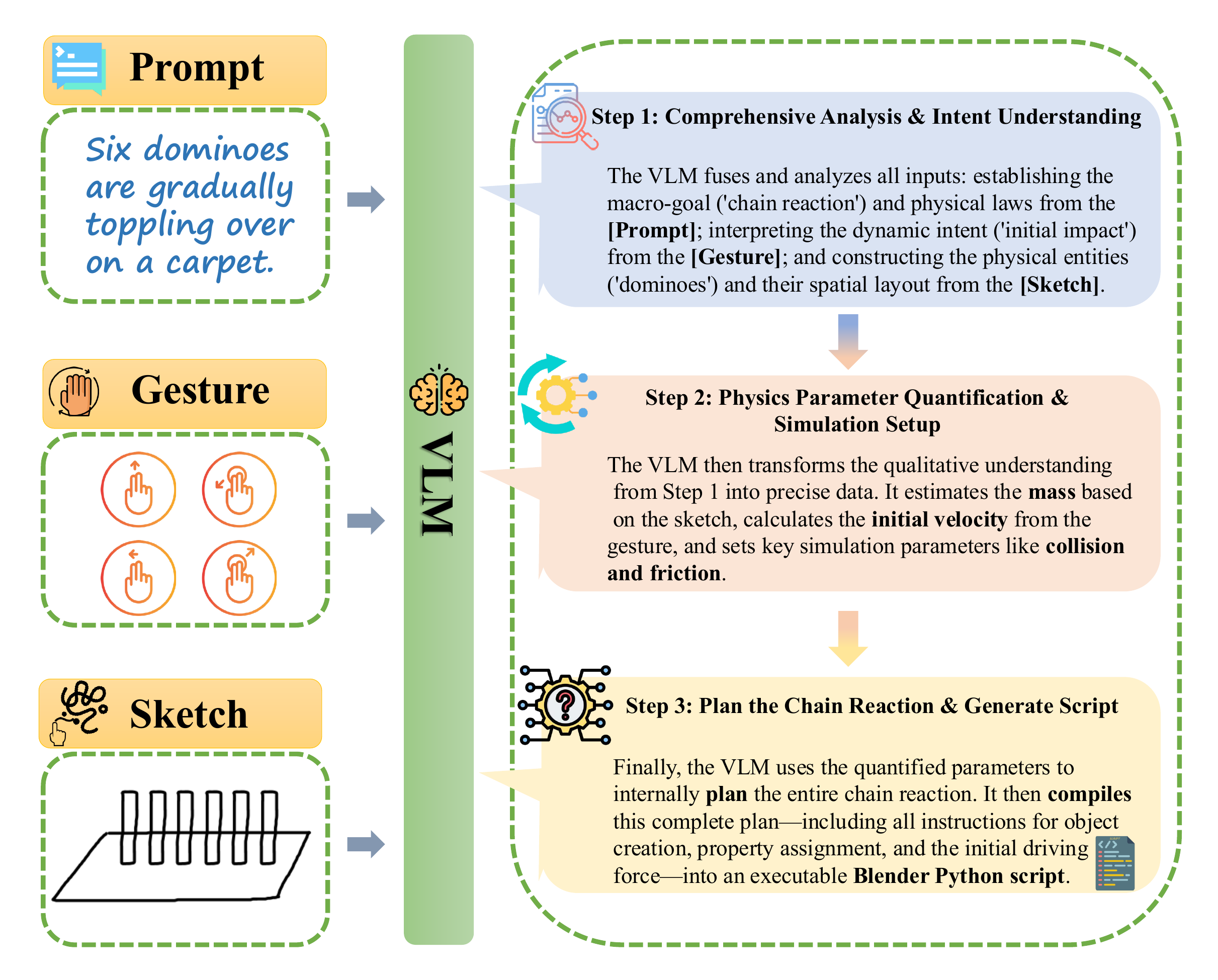} 
  \caption{\textbf{Overview of Motion Simulation.} Inputs (\textbf{Prompt}, \textbf{Gesture}, \textbf{Sketch}) are fused by a VLM to (1) parse intent and layout, (2) infer materials and mass \(m\), compute \(v_{\text{obj}}\) from the gesture with \(\alpha_{\text{material}}\), and set collision/friction, and (3) generate a Blender Python script to run the simulation, yielding physically consistent dynamics aligned with the sketch.}        
  \label{fig:cot}
\end{figure}
\section{Method}
Our method aims to convert gesture trajectory sketches into physically realistic scenes in VR environments, thereby enhancing the authenticity of interactions and expanding creative possibilities. To achieve this, as illustrated in Figure \ref{fig:framework2}, the proposed SketchPlay framework is divided into three core stages: Hand-drawn Sketch, Physics Simulation, and Scenario Synthesis. In the \textbf{Sketch Drawing and Gesture Intent Recognition} stage, we first process the input gestures using gesture recognition to extract physical motion information, including force, velocity, and direction. In the \textbf{Physically Realistic Motion Simulation} stage, this physical information is used to simulate physical behaviours such as rigid body motion, elastic deformation, and fluid dynamics. Finally, in the \textbf{Scenario Synthesis} stage, we use edge maps and depth maps as physical priors for generating physically and visually realistic scenarios. In what follows, we present these stages in detail.

\begin{figure}[t]
\centering
  \includegraphics[width=0.5\textwidth]{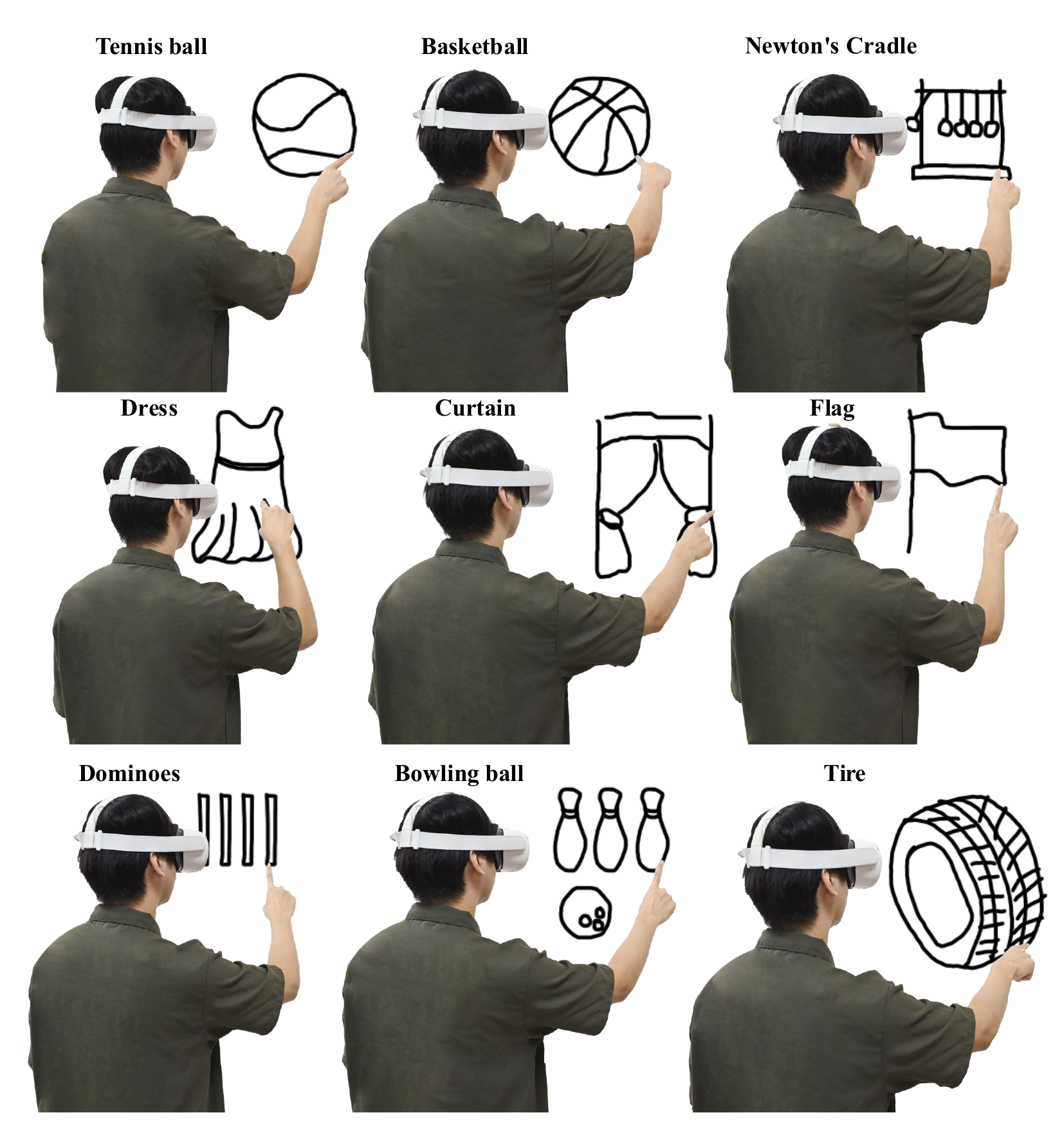} 
  \vspace{-5ex}
  \caption{Illustration of the gesture capture process in our 'Air Drawing' stage. The system uses a hand-tracking algorithm (e.g., MediaPipe\cite{zhang2020mediapipe}) to capture the keypoints of a user's hand motion in 3D space. This raw trajectory is then refined into a clean sketch, which forms the basis for the subsequent physics simulation.}
  \label{fig:gest}
  \vspace{-2ex}
\end{figure}

\subsection{Sketch Drawing and Gesture Intent Recognition}
In virtual reality, painting typically relies on tools to create artwork. However, our approach focuses solely on hand gesture trajectories to complete the drawing. To realize this vision, we divide the task into two key components: \textbf{1) Sketch Drawing}, where a rough sketch is initially drawn in mid-air, and \textbf{2) Gesture Intent Recognition}, where the hand’s trajectory is used to apply direction, velocity, and force to the sketch, thereby capturing the user’s intended movement and interaction with the artwork.
\subsubsection{Sketch Drawing}As shown in Figure \ref{fig:gest}. To capture the user's hand-drawn sketches, we use MediaPipe\cite{zhang2020mediapipe}, a hand gesture tracking algorithm that accurately captures hand keypoints. Specifically, MediaPipe\cite{zhang2020mediapipe} detects a set of 21 keypoints $\{p_1, p_2, \dots, p_{21}\}$ from the input image $I$, where each keypoint $p_i = (x_i, y_i) \in \mathbb{R}^2$ represents the 2D coordinates of a specific joint or fingertip. The process can be viewed as a mapping $f: I \rightarrow \{p_1, p_2, \dots, p_{21}\}$. However, sketches captured by MediaPipe\cite{zhang2020mediapipe} often contain curved and irregular lines, introducing noise. Therefore, we leverage the AirSketch\cite{lim2024airsketch}, which employs a controllable image diffusion model to transform these noisy data into clear and aesthetically pleasing sketches.
\subsubsection{Sketch Motion Control}To accurately capture the user's intention for sketch motion, we use MediaPipe\cite{zhang2020mediapipe} to track the movement of a single finger and obtain real-time coordinates of the keypoints. Then, we estimate the speed and direction by calculating the displacement of the keypoints between consecutive frames. Specifically, given the position of the keypoint \( p_i(t) = (x_i(t), y_i(t), z_i(t)) \) at time \( t \), the speed \( v_i(t) \) is computed using the following formula:

\[
v_i(t) = \frac{p_i(t) - p_i(t - \Delta t)}{\Delta t}
\]

The direction is represented by calculating the tangent direction between the keypoints, i.e., the difference in positions between two consecutive keypoints:

\[
\hat{v}_i(t) = \frac{p_i(t + \Delta t) - p_i(t)}{\|p_i(t + \Delta t) - p_i(t)\|}
\]

This method accurately captures the finger's speed and direction, especially in the case of curved motion, thus enabling precise capture of the user's intention for sketch movement.
\subsection{Physically Realistic Motion Simulation
}\label{sec:phys_sim}To achieve the goal of transforming sketches into physically realistic scenes, we approach the problem in two stages: \textbf{1) Physical-Aware Recognition}, where we extract essential physical cues from the sketch, and \textbf{2) Physically Realistic Scenario Simulation}, where these cues are employed to accurately simulate the movement and interactions of objects within the sketch.

\begin{figure}[t]
\centering
  \includegraphics[width=0.47\textwidth]{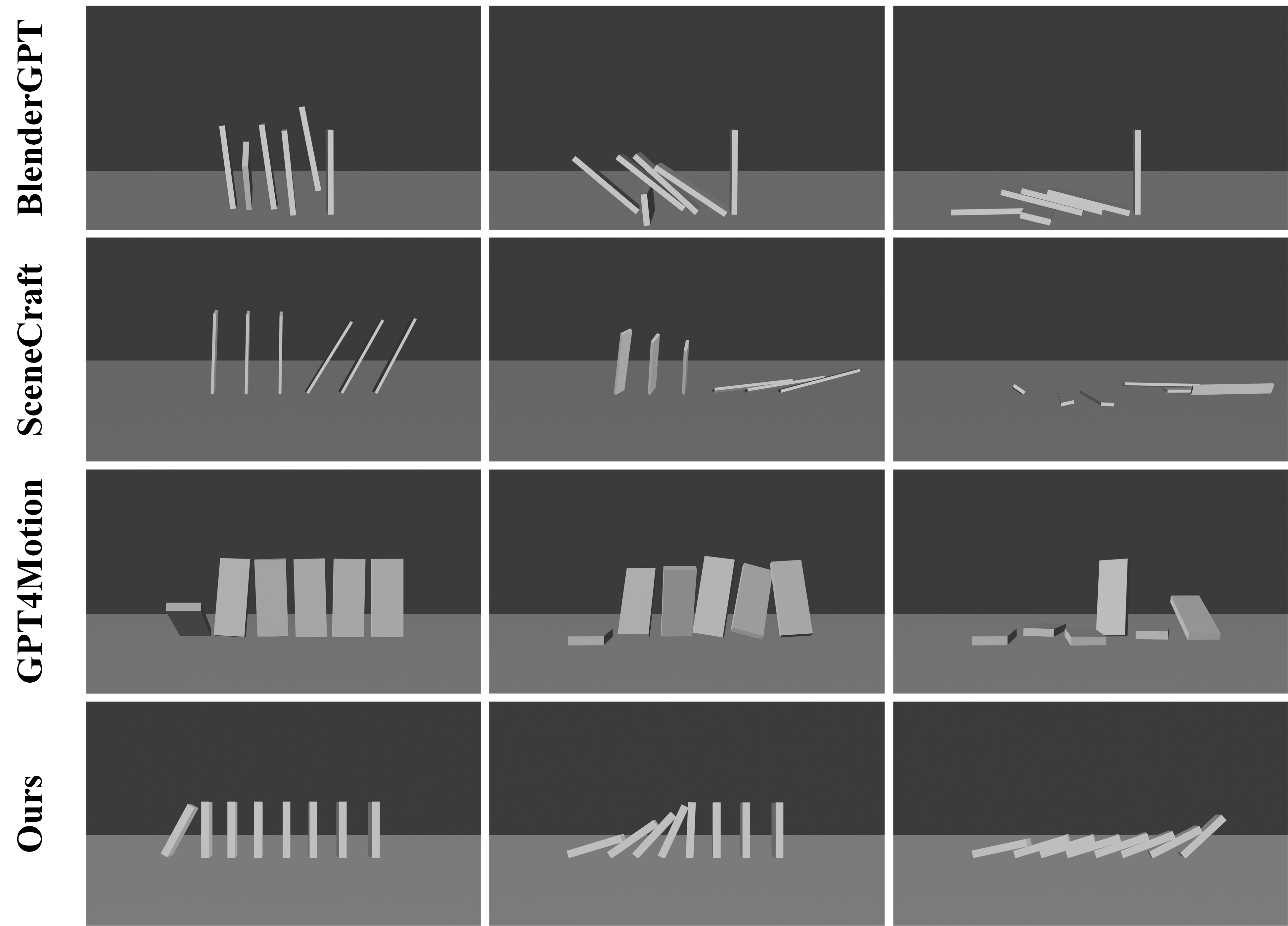} 
  \caption{Qualitative comparison of Blender simulations for the 'domino fall' task. The frames are rendered from scripts generated by four different methods. SketchPlay produces a physically plausible and coherent chain reaction, whereas scripts from baseline models result in chaotic collapses (BlenderGPT), unnatural scattering (SceneCraft), or implausible physics (GPT4Motion).}
  \label{fig:duominuo}
\end{figure}

\subsubsection{Physical-
Aware Recognition}
\begin{figure*}[t]
    \centering
    \includegraphics[width=0.9\textwidth, height=0.45\textheight]{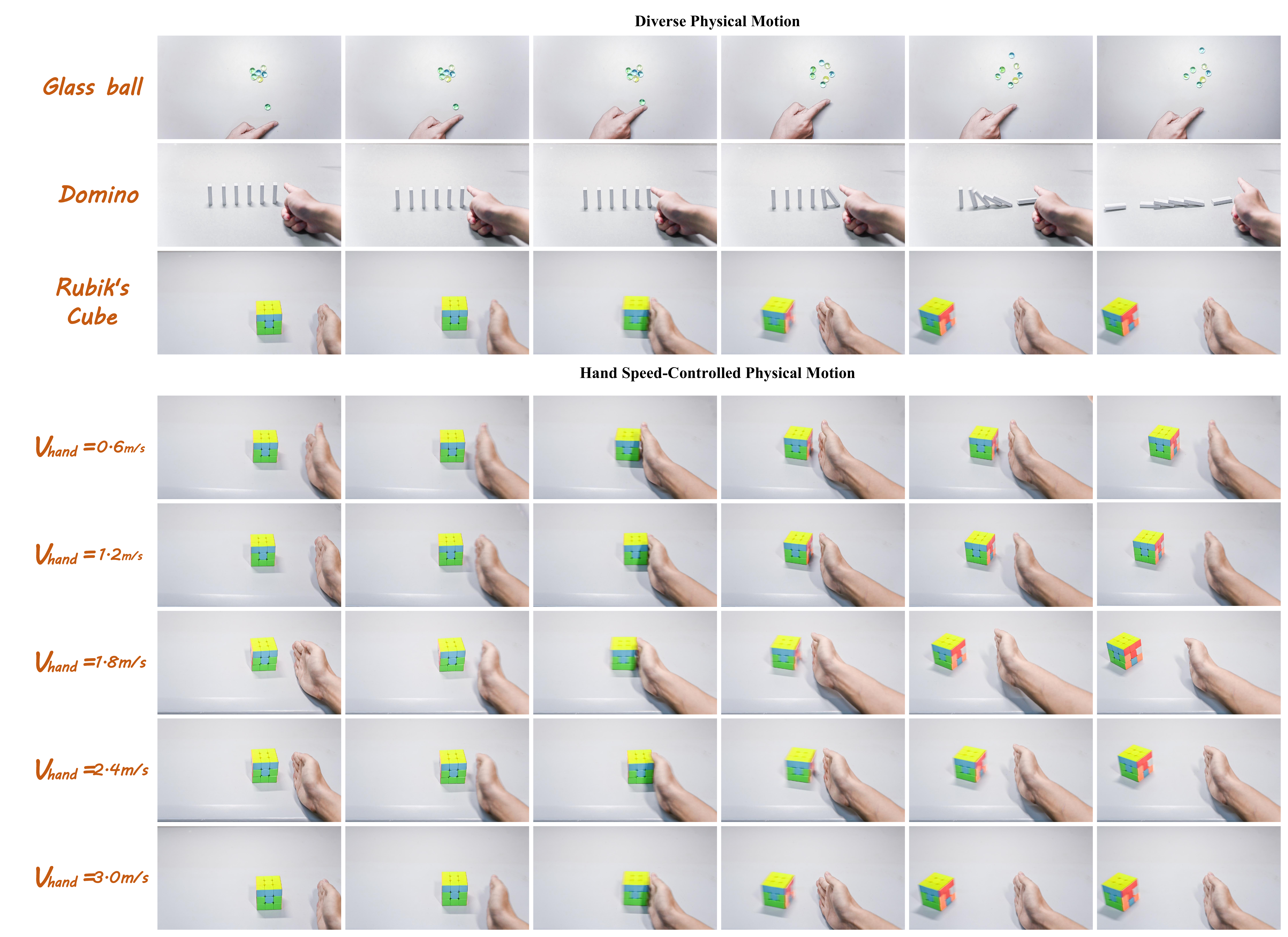}
    \vspace{-2ex}
    \caption{\textbf{Real Physical Motion}. We capture a variety of real-world physical motions, including hand-speed-controlled motions at a fine-grained level.}
    \label{fig:gt}
\end{figure*}

To accurately simulate the motion of objects in the sketch and align it with the user's intent, we first map the physical information from the user's hand motion onto the sketch. We then estimate the physical constraints that the objects must satisfy during motion, providing physically plausible cues to enable a realistic simulation. A key prerequisite for this process is the estimation of each object's material and mass based on the sketch. To achieve this, we employ a Vision-Language Model (VLM), specifically GPT-4o\cite{GPT-4o}, in a few-shot learning setting (as shown in Fig.~\ref{fig:cot}). This allows the model to infer physical properties, including material type (e.g., metal, wood, cloth) and mass (estimated from size and material) by learning from a limited number of examples or textual descriptions (see Figure \ref{fig:gt}). With this approach, the VLM can accurately predict mass and material attributes without requiring large annotated datasets.

Once the material and mass are estimated, we establish a mapping between hand motion and object response. We consider the following factors: hand velocity \( v_{\text{hand}} \), object mass \( m_{\text{obj}} \), hand mass \( m_{\text{hand}} \), and a material factor \( \alpha_{\text{material}} \) that encapsulates the object's responsiveness to external force based on material type. The velocity of the object \( v_{\text{obj}} \) is computed as:

\[
v_{\text{obj}} = v_{\text{hand}} \cdot \frac{m_{\text{hand}} + \alpha_{\text{material}} \cdot m_{\text{obj}}}{m_{\text{hand}} + m_{\text{obj}}}
\]

This formulation ensures that the object's motion is influenced not only by the hand's speed but also by its own mass and material properties. The material factor \( \alpha_{\text{material}} \) modulates the object's reaction based on physical characteristics such as hardness and elasticity. For instance, metal objects typically have a low \( \alpha_{\text{material}} \), reflecting minimal deformation in response to force, whereas cloth objects have a higher \( \alpha_{\text{material}} \), resulting in larger deformations.

By leveraging the VLM to infer material and mass, and employing the proposed motion mapping, we ensure that the simulated object motion is physically consistent and aligns with the user's intent. This approach enhances the naturalness of motion trajectories and strengthens physical plausibility in the generated sketch animations.
\subsubsection{Physically Realistic Scenario Simulation}To achieve a physically realistic simulation, we first detect and segment each object instance $p_i \in \mathbb{R}^{W \times H \times 3}$ in the sketch using Grounded-SAM\cite{ren2024grounded}, where $p_i$ represents the image of the $i$-th object. Then, we use SketchDream\cite{liu2024sketchdream} to convert the segmented objects into 3D models. Next, we utilize the \textbf{Physical-Aware Recognition} phase to extract physical cues of the objects, and use GPT-4o to generate Blender scripts for simulating the motion of the objects, thus ensuring that the motion aligns with the user's intent.

In the generated Blender script\cite{Blender}, the entire process can be divided into three stages: material assignment, physical property setup, and motion simulation. First, we assign the corresponding materials to each object based on its estimated material properties (e.g., metal, wood, cloth), which will directly apply to the 3D models converted by SketchDream\cite{liu2024sketchdream}. Then, we set the physical properties for each object, including mass \( m_{\text{obj}} \), density \( \rho_{\text{obj}} \), and rigidity \( k_{\text{obj}} \), among others. During this process, we use Blender's rigid body simulator to assign appropriate physical behavior properties to each object, ensuring that the objects' motion adheres to the laws of physics. For example, for rigid body objects, we consider the mass \( m_{\text{obj}} \), friction \( \mu_{\text{obj}} \), and elasticity \( E_{\text{obj}} \) to ensure reasonable reactions during collisions.

For elastic objects, we set the deformation response of the object based on its elastic modulus \( E \) and Poisson's ratio \( \nu \). Using stress-strain relationships, we can simulate how the object deforms under an applied force and returns to its original shape after the force is removed. This allows the object to behave in accordance with elastic materials, such as bouncing or compressing.

Additionally, for cloth objects, we set the texture, friction coefficient, and elasticity properties, allowing it to exhibit appropriate softness and dynamic response. Through these steps, the generated Blender script can accurately simulate the physical behavior of different object types, including rigid bodies, elastic objects, and cloth, ensuring that their motion in the 3D scene conforms to the laws of physics and matches the user's hand movements. This approach not only enhances the realism of the simulation but also improves the naturalness of user interactions.
\begin{figure*}[t]
  \centering
  \small
  \includegraphics[width=\textwidth]{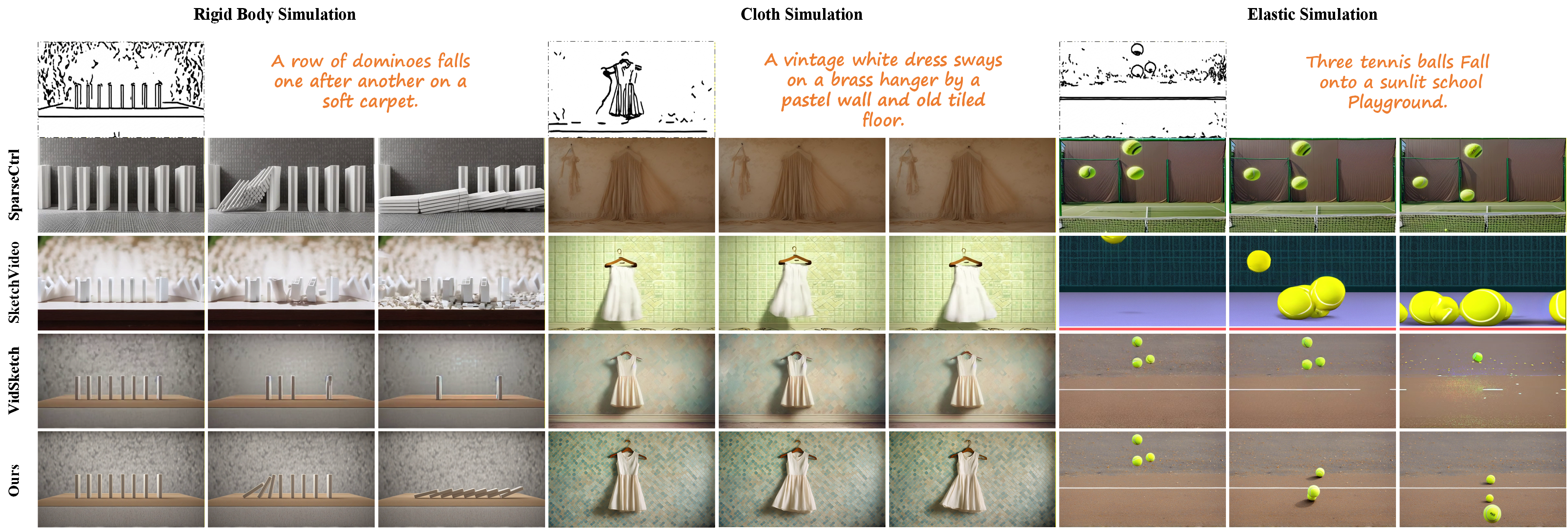} 
  \caption{Qualitative evaluation of SketchPlay against sketch-driven baselines (SparseCtrl, SketchVideo, VidSketch). Across a variety of challenging physical scenarios shown here, SketchPlay consistently generates videos with higher physical realism and fewer visual artifacts, demonstrating the effectiveness of our physics-based approach.}
  \label{fig:visual_compare}
\end{figure*}
\begin{table*}[t]
\caption{Quantitative  Evaluation. 
VBench reports automated scores for \emph{Motion Smoothness} and \emph{Imaging Quality}; Human and GPT-4o evaluations report mean opinion scores on \emph{Physical Realism} (PhysReal), \emph{Photorealism} (PhotoReal), and \emph{Semantic Alignment} (Align). 
Arrows (↑) indicate higher is better. 
Best and second-best results are marked in \textbf{bold} and \underline{underline}, respectively.}
  \label{tab:overall_results}
  \centering
  \scriptsize
  \resizebox{\textwidth}{!}{%
  \begin{tabular}{l cc ccc ccc}
    \toprule
    \multirow{2}{*}{Method} & \multicolumn{2}{c}{VBench\cite{huang2024vbench}} & \multicolumn{3}{c}{Human Evaluation\cite{chen2025physgen3d}} & \multicolumn{3}{c}{GPT-4o Evaluation\cite{chen2025physgen3d}} \\
    \cmidrule(lr){2-3} \cmidrule(lr){4-6} \cmidrule(lr){7-9}
    & Motion Smoothness $\uparrow$ & Imaging Quality $\uparrow$
    & PhysReal $\uparrow$ & PhotoReal $\uparrow$ & Align $\uparrow$
    & PhysReal $\uparrow$ & PhotoReal $\uparrow$ & Align $\uparrow$ \\
    \midrule
    SparseCtrl\cite{guo2024sparsectrl}        & 0.988               & 0.659               & 2.853              & 3.521               & 2.653              & 0.253              & 0.525               & \underline{0.613} \\
    SketchVideo\cite{liu2025sketchvideo}       & 0.988               & \underline{0.729}   & 2.670              & \underline{3.682}   & \underline{3.219}  & 0.472              & \underline{0.762}   & 0.527 \\
    VidSketch\cite{jiang2025vidsketch}         & \underline{0.990}   & 0.634               & 3.197  & 3.201               & 2.931              & 0.547  & 0.713               & 0.534 \\
    PhysGen\cite{liu2024physgen}         & 0.986   & 0.563               & \underline{3.597}  & 3.253               & 3.148              & \underline{0.562}  & 0.618               & 0.497 \\    
    Ours & \textbf{0.995}      & \textbf{0.735}      & \textbf{4.131}     & \textbf{3.857}      & \textbf{3.925}     & \textbf{0.741}     & \textbf{0.802}      & \textbf{0.720} \\
    \bottomrule
  \end{tabular}%
  }
\end{table*}
\begin{figure*}[t]
\centering
  \includegraphics[width=0.8\textwidth]{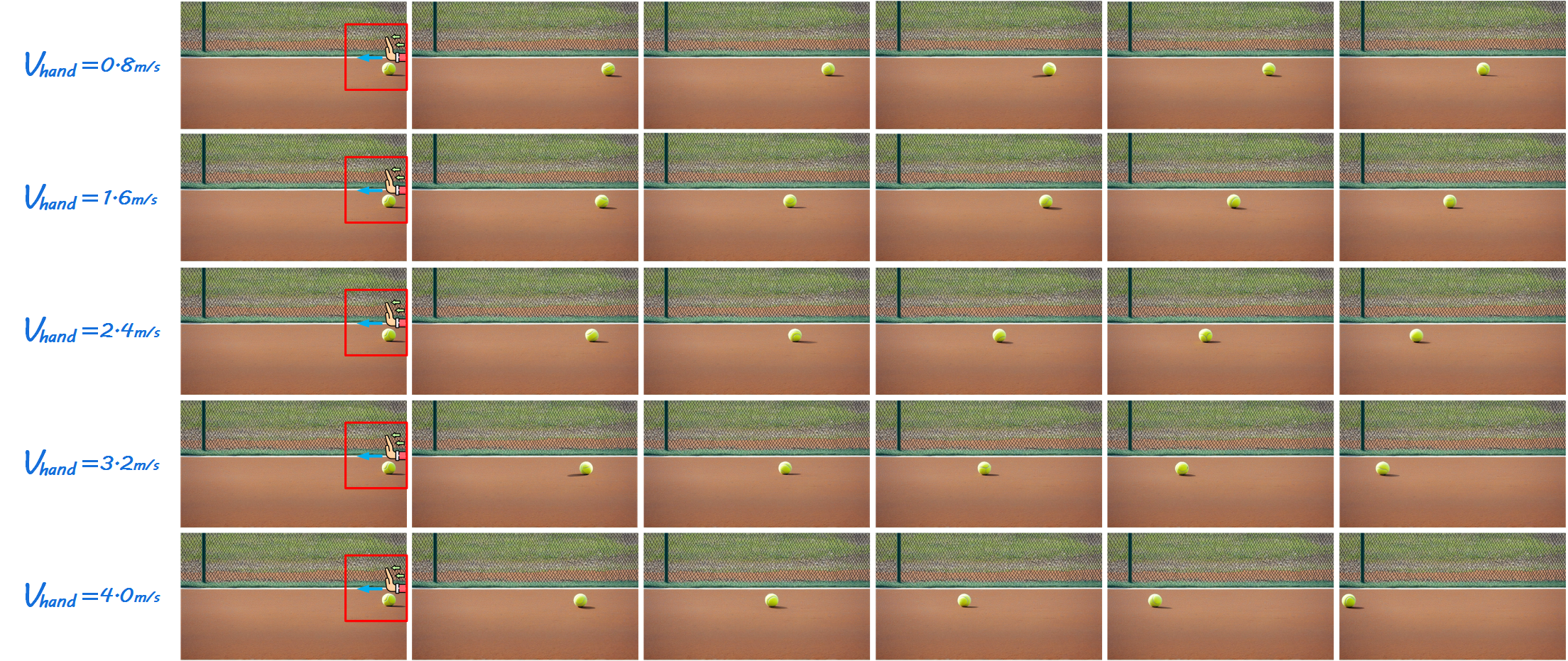} 
  \caption{Qualitative Demonstration. We demonstrate the impact of different gesture swing speeds on tennis movements.}
  \label{fig:tennis}
\end{figure*}
\begin{figure}[H]
  \centering
  \includegraphics[width=0.5\textwidth]{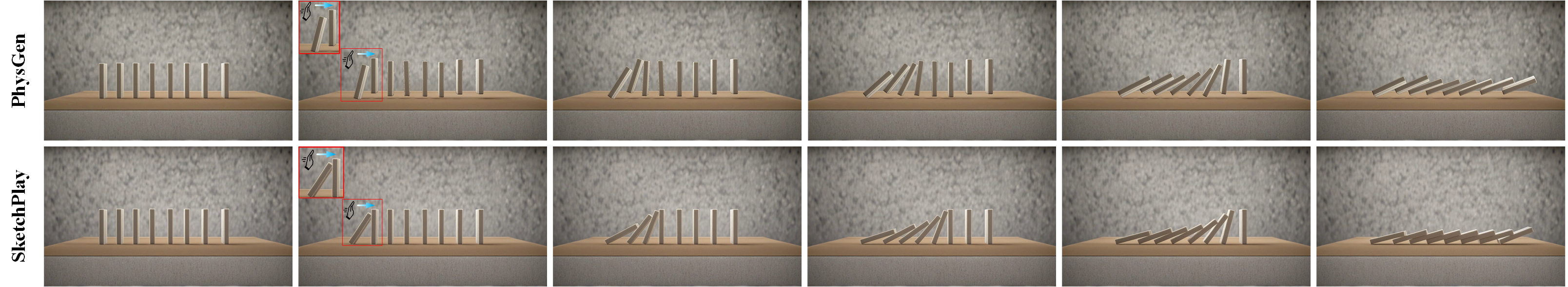} 
  \caption{\textbf{Compare With PhysGen}. PhysGen (top row) fails to maintain spatial contact, causing the dominoes to float unphysically. In contrast, SketchPlay (bottom row) correctly models the scene's geometric constraints, keeping the dominoes grounded.}
  \label{fig:duominuo-physgen}
\end{figure}
\subsection{Scenario Synthesis}
Our physics simulation stage output (\S\ref{sec:phys_sim}) is dynamically accurate but visually abstract, typically presented as sequences of textured meshes. To transform these simulation data into a realistic and interactive 4D scene, we propose a physically grounded synthesis pipeline. Directly rendering the simulation results lacks realism, while end-to-end generative models often struggle to maintain long-term physical plausibility. Our method circumvents these issues by treating the simulation results as motion priors, guiding the generation process to ensure that the final synthesis remains both visually realistic and physically consistent.

Our pipeline begins by using GPT4Motion\cite{lv2024gpt4motion} to convert the abstract simulation data into an initial single-view video, generating a visual narrative that respects the underlying physical dynamics. While the video preserves the essential motion information, it still requires further refinement due to frame-to-frame inconsistencies. We process the video using VEnhancer~\cite{he2024venhancer}, a spatio-temporal enhancement model, to increase both spatial and temporal resolution, transforming the video into a high-quality, smooth time sequence that serves as a solid foundation for subsequent steps. Next, we use the enhanced video to generate the multi-view data necessary for 4D reconstruction. For this, we employ ReCamMaster~\cite{bai2025recammaster}, a model that performs camera-controlled generative rendering from the single high-quality video. This step synthesizes a complete set of multi-view videos, $\{\hat{I}_{t,v}\}_{t=1, v=1}^{T, V}$, ensuring that the newly generated views remain physically consistent with the original motion while providing the necessary 3D scene data.

Finally, we use 4D Gaussian Splatting (4D-GS)~\cite{wu20244d} to reconstruct the final interactive 4D scene representation from the synthesized multi-view videos. 4D-GS enables efficient reconstruction of dynamically coherent scenes with temporal consistency and supports real-time rendering from novel viewpoints while maintaining motion continuity and accuracy, staying true to the original input dynamics.
\section{EXPERIMENTS AND EVALUATIONS}To comprehensively evaluate SketchPlay, we design two experiments. The first, \textbf{Script for Physical Simulation}, assesses the ability of SketchPlay to generate executable physical simulation scripts, and the results demonstrate that SketchPlay can produce scripts that are directly usable compared with strong baselines. The second, \textbf{Quality Evaluation for Generated Scenario}, evaluates the visual fidelity and physical realism of the generated scenarios, showing that SketchPlay consistently outperforms most baseline models.
 \begin{table}[tb]
  \caption{Results of the human comparison for generated Blender scripts. We compare SketchPlay against three baselines. The scores represent the percentage of times participants preferred a method's output for a given text prompt across three criteria. Higher is better.}
  \label{tab:script_quality}
  \scriptsize
  \centering
  \begin{tabu}{l c c c}
    \toprule
    Method & Text Fidelity $\uparrow$ & Composition $\uparrow$ & Aesthetics $\uparrow$ \\
    \midrule
    BlenderGPT\cite{BlenderGPT} & 12.7\% & 11.4\% & 14.5\% \\
    SceneCraft\cite{hu2024scenecraft} & 76.8\% & 83.6\% & 74.5\% \\
    GPT4Motion\cite{lv2024gpt4motion} & 65.2\% & 72.3\% & 71.0\% \\
    Ours & 80.2\% & 84.7\% & 76.4\% \\
    \bottomrule
  \end{tabu}
\end{table}


\subsection{Script for Physical Simulation}
This experiment demonstrates the high quality of Blender\cite{Blender} scripts generated by SketchPlay through a comparison with other baseline methods.\\
\textbf{Baselines and Metrics. }To evaluate the performance of our method SketchPlay, we compare it against three baselines: BlenderGPT\cite{BlenderGPT}, SceneCraft\cite{hu2024scenecraft}, and GPT4Motion\cite{lv2024gpt4motion}. We follow the evaluation metrics used in SceneCraft to comprehensively assess the quality of the generated Blender scripts. \textit{Text fidelity} is used to evaluate how closely the generated scene aligns with the given textual description and determine the accuracy with which the scene reflects the described elements. \textit{Composition and constraint agreement} is used to assess the object relations and spatial arrangements in the scene generated by the Blender script. \textit{Aesthetics} is used to evaluate the overall visual quality of the generated scene, considering factors such as visual appeal, realism, and how well the scene presents a cohesive and aesthetically pleasing composition.\\
\textbf{Evaluation Results. }
We conducted a qualitative human study to obtain the results for this comparison. We recruited multiple participants with experience in 3D modeling and utilized 6 diverse text prompts. For each prompt, participants were shown the rendered outputs from all four methods in a randomized, side-by-side comparison and were asked to select which output best fulfilled the prompt. The percentage scores, representing this preference rate, are reported in Table~\ref{tab:script_quality}.The data shows that our method, SketchPlay, significantly outperforms all baselines across every evaluated dimension. Its advantages are most pronounced in Composition (\textbf{84.7\%}) and Aesthetics (\textbf{76.4\%}), which proves that our framework is capable of generating more logically correct and visually superior physical scenarios. This is further corroborated by the qualitative comparison in Figure~\ref{fig:duominuo}. SketchPlay successfully generates a stable and coherent chain reaction, whereas the baseline models exhibit significant logical or physical flaws.
\subsection{Quality Evaluation for Generated Scenario}
This experiment aims to evaluate the quality of videos generated by SketchPlay against other baseline methods.\\
\textbf{Baselines and Metrics. }Figure \ref{fig:visual_compare} , \ref{fig:tennis} and \ref{fig:duominuo-physgen} presents a qualitative comparison of three physical simulations, benchmarking our method against two categories of baselines: gesture-driven and sketch-driven video generation. For the gesture-driven baseline, we compare against PhysGen\cite{liu2024physgen}, which generates videos through rigid body simulations guided by force inputs. For sketch-driven synthesis, we provide a qualitative comparison of video generation quality against three state-of-the-art models: SparseCtrl\cite{guo2024sparsectrl}, SketchVideo\cite{liu2025sketchvideo}, and VidSketch\cite{jiang2025vidsketch}. We note that VidSketch additionally requires a reference image from the first frame to condition its generation.

The absence of ground-truth data for our generative task precludes the use of conventional, pixel-based metrics (e.g., PSNR, SSIM, LPIPS) or distribution-based metrics (e.g., FID, FVD). To address this limitation, we adopt the evaluation protocol from recent works like PhysGen3D\cite{chen2025physgen3d} and VidSketch\cite{jiang2025vidsketch}, performing two complementary sets of assessments. First, we employ the VBench\cite{huang2024vbench} benchmark suite for an automated evaluation of Motion Smoothness and Imaging Quality. Motion Smoothness quantifies the temporal consistency and fluidity of object dynamics, while Imaging Quality assesses the perceptual fidelity by detecting visual artifacts or distortions.Second, we conduct both human evaluations and leverage the advanced multimodal capabilities of GPT-4o\cite{GPT-4o} to assess three critical qualitative metrics: Physical Realism (PhysReal), Photorealism (Photoreal), and Semantic Consistency (Align). Specifically, PhysReal evaluates the adherence of the generated video to fundamental physical laws, such as gravity and momentum. Photoreal measures the overall visual fidelity, focusing on artifacts, temporal coherence, and the plausibility of lighting and shadows. Finally, Align quantifies the semantic consistency between the generated video content and the user's input control conditions.\\
\textbf{Evaluation Results. }We present a comprehensive quantitative analysis in Table \ref{tab:overall_results}, summarizing the performance of SketchPlay against sketch-driven baseline methods across both automated and qualitative metrics, with results consistently demonstrating the superiority of our proposed method. This is further substantiated by the qualitative comparisons in Figure \ref{fig:visual_compare}. In the VBench\cite{huang2024vbench} evaluation, SketchPlay achieves state-of-the-art scores for both Motion Smoothness (0.995) and Imaging Quality (0.735), indicating that our physics-based pipeline generates videos with superior temporal consistency and fewer perceptual artifacts. Furthermore, this superiority extends emphatically to the qualitative evaluations conducted by both humans and GPT-4o. In the human study, our method significantly outperforms all baseline models across Physical Realism, Photorealism, and Semantic Consistency. Notably, its PhysReal score of 4.131 marks a substantial improvement over the next best method, VidSketch (3.197), empirically confirming that our physics-informed approach generates dynamics perceived as significantly more plausible. This quantitative lead is visually corroborated by the comparisons in Figure \ref{fig:visual_compare}, where SketchPlay successfully renders a coherent domino fall and realistic ball bounces, while baselines often exhibit physically implausible motions.
In contrast to the gesture-driven baseline PhysGen~\cite{liu2024physgen}, which generates rigid-body simulations from force inputs, it is built on a 2D simulator and therefore cannot model true 3D spatial relationships (e.g., depth ordering and out-of-plane contacts), leading to less favorable results in scenarios requiring intricate 3D object interactions(see Figure\ref{fig:duominuo-physgen}).
These overall findings are reinforced by the GPT-4o evaluation, where SketchPlay again secures the top position across all three qualitative metrics. The strong agreement between automated, human, and AI-based assessments validates the effectiveness of SketchPlay in generating high-quality, physically realistic video content that robustly aligns with user intent.

\section{Conclusion and Future Work}In this paper, we introduce SketchPlay, a novel VR interaction framework that lowers the technical barrier to VR content creation by seamlessly transforming a user's freehand gestures and air-drawn sketches into physically and visually realistic 4D scenes. Our core contribution lies in a unique method that combines the dynamic information of gestures (e.g., velocity, direction) with physical properties inferred by a VLM (e.g., mass, material) to automatically generate an executable physics simulation. Finally, through an advanced synthesis pipeline, the simulation results are elevated into photorealistic multi-view videos and reconstructed into a complete 4D scene. Experiments demonstrate that SketchPlay significantly outperforms existing methods in generating both physical realism and high-quality visual effects. Future work will focus on expanding the scope to more complex physical phenomena, such as fluid dynamics, and exploring the implementation of an immersive creative environment that offers real-time interactive feedback and supports multi-user collaboration, further unlocking the creative potential of users in fields like education, art, and storytelling.

\bibliographystyle{abbrv-doi}

\bibliography{template}
\end{document}